\documentstyle{aa}     
\begin{document}
 
   \thesaurus{08        
             (02.18.7;  
              08.03.1;  
              08.03.4)  
             }
   \title{Modeling of C stars with core/mantle grains: Amorphous carbon + 
SiC}
  
   \author{S. Lorenz-Martins\inst{1}, F.X. de Ara\'ujo\inst{2}, S.J. Codina 
Landaberry\inst{2}, W.G. de Almeida\inst{2} \&  R.V. de Nader\inst{1}}

   \institute{Observatorio do Valongo/UFRJ,
              Ladeira Pedro Antonio 43,
              20080-090 - Rio de Janeiro - Brazil
   \and
              Observatorio Nacional/MCT, 
              Rua Gal. Jose Cristino 77,
              20921-400 - Rio de Janeiro - Brazil}
 
   \date{Received 4 August 2000 / Accepted 9 November 2000}

\authorrunning{S.Lorenz-Martins et al.}
\titlerunning{Modeling of C stars with core/mantle grains} 
 
   \offprints{S. Lorenz-Martins}
 
 \maketitle
 
 \input epsf
 
   \begin{abstract}
A set of 45 dust envelopes of carbon stars has been modeled.  Among them, 34 
were  selected according to their dust envelope class (as suggested by Sloan, 
Little-Marenin \& Price, 1998) and 11 are extreme carbon stars. The models were 
performed using a code that describes the radiative transfer in dust envelopes 
considering core/mantle grains composed by an $\alpha$-SiC core and an 
amorphous carbon (A.C.) mantle.  In addition, we have also computed models with 
a code that considers two kinds of grains - $\alpha$-SiC and A.C. - 
simultaneously.  Core-mantle grains seem to fit dust envelopes of evolved carbon 
stars, while two homogeneous grains are more able to reproduce thinner dust 
envelopes.  Our results suggest that there exists an evolution of dust grains in 
the carbon star sequence. In the beginning of the sequence, grains are mainly 
composed of SiC and amorphous carbon; with dust envelope evolution, carbon 
grains are coated in SiC. This phenomena could perhaps explain the small 
quantity of SiC grains observed in the interstellar medium. However, in 
this work we consider only $\alpha$-SiC grains, and the inclusion of $\beta$-SiC 
grains can perhaps change some of there results.

      \keywords{Stars:carbon - circumstellar matter - radiative transfer - 
                         SiC grain }
   \end{abstract}
 
%
\section{Introduction}

Asymptotic giant branch (AGB) stars are often surrounded by circumstellar dust 
shells.  The chemical composition of these media reflects that of the stellar 
photosphere.  Thus, carbon-rich grains, such as amorphous carbon (A.C.),  
are one of the expected components of the dust envelopes around carbon 
stars.  In addition, almost all these stars show an emission feature around 
11.3$\mu$m due to Silicon carbide (SiC) grains which are also condensed there.  
The existence of SiC grains in the atmospheres of carbon stars was predicted at 
first on the basis of chemical equilibrium calculations by Gilman (1969).  
This prediction was supported by the observations of Hackwell (1972), Treffers 
\& Cohen (1974) and Forrest et al. (1975). Nowadays infrared satellites provide 
several characteristic features of these grains.

After the observations with IRAS satellite, several works have dealt with the  
classification of dust envelopes around carbon stars.  Little-Marenin 
et al. (1987) have discovered 176 new carbon stars using the feature at 
11.3$\mu$m as a selection factor.  Willems (1987) has analyzed 304 such 
objects, suggesting a SiC-index to the stars which present the emission feature 
varying between 11.2--11.6$\mu$m.  Papoular (1988) has classified carbon dust 
envelopes using the 11.3$\mu$m feature and a secondary feature at 8.6$\mu$m.  
The sample of Chan \& Kwok (1990) was composed of 356 objects which were 
classified in two distinct classes. According to the authors, the difference 
between the classes is due to an evolution of $\alpha$-SiC and 
$\beta$-SiC particles. More recently Sloan, Little-Marenin \& Price (1998, 
hereafter SLMP) proposed a classification of 89 carbon-rich stars in 6 
different types based on their infrared emissions.  All stars of their sample 
show the SiC feature at 11.3$\mu$m.

In general, differences between each class can be interpreted as an evolution 
of the dust envelope itself due to an increasing amount of grains; consequently,  
optical depth also increases, affecting radiative transfer and the 
emission feature. On the other hand, the resonance features of small SiC grains 
are very sensitive to size, morphology and chemical composition of 
impurities in the surrounding medium (Bohen \& Huffman 1998).  This fact 
suggests that the variety of emission features assigned to SiC grains is 
probably related to the formation process of the dust in circumstellar 
envelopes, reflecting the physical and chemical conditions.  The classification 
criteria cited above are useful, but if one is interested in a deeper 
insight into the nature of the circumstellar dust grains, it is necessary to 
solve the radiative transfer problem, and to reproduce the features seen in the 
mid-IR LRS together with the  overall behavior of the spectral energy
distribution.

Several authors have calculated the radiative transfer in circumstellar dust 
shells (CDS) of carbon stars (Chan \& Kwok 1990; Lorenz-Martins \& Lef\`evre 
1993, 1994; Groenewegen 1995; Bagnulo, 1996; Bagnulo, Doyle and Andretta 1998).
Lorenz-Martins \& Lef\`evre (1994) have employed the Monte Carlo method for 
solving the radiative transfer for two species simultaneously. SiC grains were 
supposed to form closer to the star than graphite grains (McCabe 1982). 
Correlations between SiC/A.C. ratio and extinction opacity as well as  
SiC/A.C. ratio and period of luminosity were found.  These correlations indicate 
that the quantity of SiC grains relative to amorphous carbon grains decreases 
with carbon star evolution. Mass loss from cool stars is the 
major source of refractory grains in the interstellar medium.  Carbon stars 
provide carboneous material and extreme carbon stars, with their large mass-loss 
rate, are the main contributors.  A study by Whittet et al. (1990) shows that 
there is a low fraction ($<$ 5\%) of Si in the form of SiC in the interstellar 
medium. The enrichment rate for a kind of grain depends on the composition of 
the dust envelopes of stars which have high mass-loss rate. Lorenz-Martins \& 
Lef\`evre (1994) also suggested that SiC grains are the minor component in 
carbon star envelopes. In addition, stars which have low SiC/A.C. ratios 
(0.01--0.06) have also high mass-loss rates (1.5.$10^{-4}$ -- IRC+10216 --- 
6.1.$10^{-6}$ $M_{\odot}/yr$ -- IRAS 15194-5115).  These results can 
perhaps explain the small quantity of SiC grains in the interstellar medium.  
Alternatively, the weakness of the SiC absorption  could be due to the fact that 
SiC particles are embedded in thick carbon mantles. In fact, Kozasa et al. 
(1996) have shown that the nucleation of SiC precedes that of carbon grains and 
may lead to the formation of dust grains consisting of a SiC core and a carbon 
mantle.  Moreover they proposed that such core/mantle  grains are the most 
reasonable candidates to reproduce the feature seen around 11.3$\mu$m. They also  
suggested that radiative transfer calculations should be used in order to verify 
this model.

The main purpose of the present work is to present dust envelope models     
considering core/mantle  grains composed by an $\alpha$-SiC core and 
an A.C. mantle. We also compare the data with models consisting of $\alpha$-SiC  
and A.C. homogeneous grains. Two samples of stars have been 
considered. The first one contains objects classified by SLMP. They were 
analyzed aiming to verify the classification proposed by the authors.  The 
second sample contains 11 extreme carbon stars, which have thicker dust 
envelopes and a higher mass loss rate. The utilization of core/mantle  
grains is satisfactory to describe some stars but most of them are better 
reproduced using a two homogeneous grains model.  

\section{Core/mantle grains and the method}

Besides carbon grains of different structures (e.g., graphite, amorphous   
carbon), SiC is the most important species for late-type 
carbon stars. SiC is one of the most refractory materials that may condense 
under the conditions of a carbon-rich chemistry.  In such stars, almost all 
oxygen is chemically blocked in the CO molecule and, among the more abundant 
chemically active elements, carbon and SiC are the possible condensates. An 
important problem in the formation of SiC under the conditions present in 
circumstellar shells is that the abundance of SiC molecule is quite low.  
Beck (1992) has considered non-equilibrium effects and has shown that solid 
SiC can be stable against evaporation at temperatures below 1400K.  He also 
suggested that SiC may form on the surface of preexisting carbon grains instead 
of being the primary condensate at very high temperatures. On the other hand, 
McCabe (1982) has shown that SiC particles can be formed at high temperatures 
due to the greenhouse effect. Following this suggestion, Kozasa et al. (1996) 
have demonstrated that the nucleation of SiC grains always precedes that of 
carbon grains when the non-LTE effect, i.e., the difference between the 
temperature of gas and small clusters, is taken into account. 

We have considered this latest suggestion and modeled 45 envelopes of carbon 
stars using core/mantle  grains consisting of a $\alpha$-SiC core 
and a A.C. mantle. The method employed is an improved version of that described 
by Lorenz-Martins \& Ara\'ujo (1997). We have modified the code to include a 
new option about grain properties. The absorption and scattering efficiencies, 
as well as the albedo, were calculated using the Mie theory for core/mantle  
grains (e.g. Bohren \& Huffman, 1984; Hoyle \& Wickramasinghe, 1991) and optical 
constants (or dielectric functions) tabulated in the literature. The optical 
constants which we have used are the ones determined by Pegouri\'e (1988) for  
$\alpha$-SiC, and by Rouleau \& Martin (1991) for amorphous carbon.   

The propagation of stellar and grain radiative energy is simulated photon by 
photon following a Monte Carlo scheme.  For each interaction between a 
``photon'' and a grain, a fraction of the energy is stored (absorption) and 
the remaining part is scattered according to the scattering diagram.  The 
stellar radiation leads to an initial distribution of dust temperature and the 
thermal radiation from grains is simulated, giving after several iterations the 
equilibrium temperature. Computations give the spectral repartition of the 
total flux and of its different components (direct, scattered, emitted), and the  
temperature law for the grains. For more details see Lorenz-Martins \& Lef\`evre 
(1993, 1994) and Lorenz-Martins \& Ara\'ujo (1997).

\section{The sample}

SLMP have studied 89 carbon-rich stars and organized the dust emission in 
several classes. {\it Red} class contains only 3 stars which present a 
11.3$\mu$m feature and a strong dust continuum. {\it SiC} class is the most 
numerous, with 40 objects.  They have the 11.3$\mu$m feature and a weak dust 
continuum.  In the {\it SiC+} they put 32 stars which show two features: the 
11.3$\mu$m one and a weak feature at 8-9$\mu$m. These objects also have a 
weak dust continuum. The {\it SiC++} class contains 6 stars with comparable 
11.3$\mu$m and 8-9$\mu$m features. The five stars  in the {\it Broad 1} class 
have an unusual 11.3$\mu$m feature profile with short-wavelength excess.  
Finally, the {\it Broad 2} class contains only 3 stars which present an unusual 
11.3$\mu$m feature profile with long-wavelength excess.  

In Table 1 we present our sample of 34 stars taken from SLMP. We restrict our 
study to the objects with IR fluxes published in the literature. Fortunately, we 
have been able to obtain data from stars belonging to all different classes. 
Table 1 lists the IRAS number (column 1) followed by the usual name (column 2).  
Spectral type and variability class are listed in columns 3 and 4 respectively.  
Column 5 gives the period and column 6 shows the envelope class attributed by 
SLMP. Some stars have a doubtful classification; they are designated by a colon. 
Finally the photometry used in the fit of the models is presented in the last 
column.  Table 2 list the extreme carbon stars sample. The columns are analogous 
to those of Table 1.  The stars analyzed are variables and self-consistent 
models for them require data from similar phase of luminosity. However this is 
difficult due to the scarcity of observations.  We have worked with the 
photometry available in the literature and considered the phase whenever 
possible. In order to minimize the uncertainties, we have used average LRS 
(IRAS, 1986) spectra for SiC emission.  The 12, 25, 60, 100 $\mu$m fluxes were 
taken from Gezari et al. (1987) and SIMBAD.

  \begin{table*}
      \begin{flushleft}
       \caption{Sample of SLMP's stars}     
      \begin{tabular}{lllllll}
      \hline\noalign{\smallskip}
IRAS & Name  & Sp.Type & {Var.}  & {Period} & Env.class$^i$ & Phot.\\
      \noalign{\smallskip}   
      \hline\noalign{\smallskip} 
00172+4425   & VX And    & C4,5J   & SRa    & 369$^a$  & SiC++   & 9\\
01246-3248   & R Scl     & C6,5    & SRb    & 370$^a$  & SiC++   & 1,7\\
02270-2619   & R For     & C4,3e   & Mira   & 388$^c$  & SiC     & 8\\
03075+5742   & C* 131    & C4,5J   & Lb     & ---      & N:      & 9\\
03374+6229   & U Cam     & C6,4    & SRb    & ---      & SiC+    & 9\\
04459+6804   & ST Cam    & C5,4    & SRb    & 300$^a$  & SiC+:   & 9\\
04483+2826   & TT Tau    & C7,4    & SRb    & 166$^a$  & SiC+:   & 9\\      
04573-1452   & R Lep     & C7,4e   & Mira   & 427$^a$  & SiC     & 2,8\\
05028+0106   & W Ori     & C5,4    & SRb    & 212$^a$  & SiC+    & 2,9\\
05418-4628   & W Pic     & C       & Lb     & ---      & SiC++:  & 7\\
05426+2040   & Y Tau     & C6,4    & SRb    & 241$^a$  & SiC     & 7\\ 
05576+3940   & AZ Aur    & C8      & Mira   & 416$^a$  & Br2     & 9\\
06225+1445   & BL Ori    & C6,3    & Lb     & ---      & Br2     & 3,9\\
06331+3829   & UU Aur    & C7,4    & SRb    & 23$a$    & SiC     & 2,9\\
06529+0626   & CL Mon    & C6,3e   & Mira   & 497$^a$  & SiC     & 9 \\
07045-0728   & RY Mon    & C5,5    & SRa    & 278$^a$  & SiC+    & 9,12\\
07057-1150   & W CMa     & C6,3    & Lb     & ---      & SiC+:   & 7,9\\
07065-7256   & R Vol     & Ce      & Mira   & 454$^a$  & SiC     & 5\\ 
08538+2002   & T Cnc     & C5,5    & SRb    & 482$^a$  & SiC++   & 2,3\\
09452+1330   & IRC+10216 & C9,4    & Mira   & 649$^c$  & Red     & 11\\
10329-3918   & U Ant     & C5,3    & Lb     & ---      & SiC+:   & 7\\
10350-1307   & U Hya     & C5,3    & Lb     & ---      & SiC     & 2,5,9\\
10491-2059   & V Hya     & C6,5    & SRa    & 531$^a$  & Red     & 2,5,9\\
12226+0102   & SS Vir    & C6,3    & SRa    & 364$^a$  & Br1     & 9\\
12427+4542   & Y CVn     & C5,5J   & SRb    & 158$^a$  & SiC+:   & 2,9,11\\
12447+0425   & RU Vir    & C8,1e   & Mira   & 433$^a$  & SiC     & 5\\
12544+6615   & RY Dra    & C4,5J   & SRb    & 200$^a$  & SiC+    & 2,9\\
15094-6953   & X TrA     & C5,5    & Lb     & ---      & SiC+    & 5,11,12\\
18306+3657   & T Lyr     & C6,5J   & Lb     & ---      & SiC++   & 9\\
19017-0545   & V Aql     & C6,4    & SRb    & 353$^a$  & SiC+    & 5,9\\
19555+4407   & AX Cyg    & C4,5    & Lb     & ---      & SiC+:   & 9\\
21032-0024   & RV Aqr    & C6,3e   & Mira   & 454$^a$  & SiC     & 5\\
21399+3516   & V460Cyg   & C6,4    & SRb    & 263$^a$  & SiC+:   & 2,9\\
23587+6004   & WZ Cas    & C9,2J   & SRb    & 186$^a$  & Br1     & 3,10\\
\noalign{\smallskip} \hline
\end {tabular}
\end{flushleft}
(1) Bagnulo et al. (1998) (2) Bergeat et al. (1976); (3) Bergeat \& Lunel 
(1980); (4) Epchtein et al. (1990); (5) Fouqu\'e et al. (1992); (6) Jones et 
al. (1990); (7) Kerschbaum, Lazaro\& Habison (1996); (8) Le Bertre (1992); 
(9) Noguchi et al.(1981); (10) Noguchi \& Akiba (1986); (11) Nyman et al. 
(1992); (12) Walker (1976). Periodos: ($a$) Kholopov et al. (1987), 
($b$) Jones et al. (1990), ($c$) Le Bertre (1992).  ($i$) SLMP

\end{table*}

  \begin{table}
      \begin{flushleft}
      \caption{Sample with evolved stars}
      \begin{tabular}{llllll}
      \hline\noalign{\smallskip}
IRAS & Name  & Sp.Type & {Period} & Phot.\\
      \noalign{\smallskip}   
      \hline\noalign{\smallskip} 
05377+1346   & AFGL 799  &\ \ C8,4        &\ 372$^c$    & 1,4\\
05405+3240   & AFGL 809  &\ \ C           &\ 780$^b$     & 5\\
06012+0726   & AFGL 865  &\ \ ?           &\ 696$^c$    & 6\\
06291+4319   & AFGL 954  &\ \ C           &\ ---        & 5\\
06342+0328   & AFGL 971  &\ \ C           &\ 653$^c$    & 6\\
07098-2112   & AFGL 1085 &\ \ N           &\ 725$^c$    & 6\\
08088-3243   & AFGL 1235 &\ \ C           &\ 571$^c$    & 2\\
15082-4808   & AFGL 4211 &\ \ ?           &\ ---        & 3\\ 
19594+4047   & AFGL 2494 &\ \ C           &\ 783$^b$    & 5\\
20570+2714   & AFGL 2686 &\ \ C8,5        &\ 750$^b$    & 5\\
23257+1038   & AFGL 3099 &\ \ C           &\ 484$^c$    & 5\\
\noalign{\smallskip} \hline
\end {tabular}
\end{flushleft}
(1) Cohen (1984); (2) Epchtein et al. (1990); (3) Fouqu\'e et al. (1992); 
(4) Gehrz \& Hackwell (1976); (5) Jones et al. (1990); (6) Le Bertre (1992); 
Periods: ($a$) Kholopov et al. (1987), ($b$) Jones et al. (1990), 
($c$) Le Bertre (1992).
\end{table}


\section{Results}

Table 3 and Table 4 present the results of the best models to our first and 
second samples, respectively.  These results were obtained considering a 
core/mantle  grain. IRAS number (column 1) is followed by the temperature 
of the central star (T${\rm eff}$ in K) in the second column.  Third and fourth 
columns present inner (R$_1$ in R$_{\star}$) and outer (R$_2$ in R$_{\star}$) 
envelope radii, respectively. The dimension of the $\alpha$-SiC core 
(c$_{\rm SiC}$ in \AA) of the grains is given in the fifth column,  followed by 
the dimension of the amorphous carbon mantle (m$_{\rm A.C.}$ in \AA) in the 
sixth column. Finally, the optical depth ($\tau$) at 1$\mu$m and the abundance 
ratio between $\alpha$-SiC and amorphous carbon grains are given, respectively, 
in columns 7 and 8. In Table 3 we have added one last column with the SLMP's 
envelope classes.  

In addition, we have calculated models for the stars in the sample using a code 
with two homogeneous grains.  Some stars of this sample were analyzed in 
Lorenz-Martins \& L\'efevre (1994), where the authors describe the method.  The 
differences between parameters of both codes are the size of the grains and the 
way in which the SiC/A.C. ratios were calculated. In the core/mantle models, we 
consider the core (c$_{\rm SiC}$) and mantle (m$_{\rm A.C.}$) size.  SiC/A.C. 
ratios in this method are obtained by mass, based on the mantle and core size, 
and we find the corresponding value as obtained in the two homogeneous grains 
model. Others parameters are obtained the same way as in the two homogeneous 
grains 
code.

The features seen in the mid-IR LRS must be reproduced and the properties of 
complete CDS must be determined simultaneously.  In order to fit the dust 
emission, we have calculated grids of about fifty models for each star, and we 
inspect visually the model which best reproduces the complete CDS.  We pay 
special attention to the LRS feature and accept errors of about 15 per cent in 
all parameters (such as grain size, effective temperature, optical depth...) to 
fit this emission feature; in fact it is this feature that defines the best 
model.

%
   \begin{table*}[h]
       \caption{Results of first sample}
       \begin{flushleft}
      \begin{tabular}{lllcccccl}     
      \hline\noalign{\smallskip} 
Stars   & $T_{\rm eff} (K)$  & R$_{\rm 1}$ (R$_{\star})$  & R$_{\rm 2}$ 
(R$_{\star}$) & $c_{SiC}$ (\AA) & $m_{A.C.}$ (\AA) & $\tau$ & SiC/A.C. & Env. 
Class.\\
      \noalign{\smallskip}            
      \hline\noalign{\smallskip} 
00172+4425   & 2200   & 5.0  & 1000  & 453  & 1000   & 0.02  & 0.10 & SiC++\\
01246-3248   & 2400   & 5.0  & 1000  & 500  & 1200   & 0.10  & 0.08 & SiC++\\
02270-2619   & 2500   & 5.0  & 5000  & 407  & 1050   & 7.00  & 0.06 & SiC\\
03075+5742   & 2400   & 5.0  & \ 800 & 453  & 1000   & 0.01  & 0.10 & N:\\
03374+6229   & 2650   & 4.6  & 1000  & 500  & 1200   & 0.50  & 0.08 & SiC+\\
04459+6804   & 2700   & 4.8  & 1000  & 500  & \ 850  & 0.03  & 0.26 & SiC+:\\
04483+2826   & 2650   & 3.0  & 1000  & 236  & \ 400  & 0.02  & 0.26 & SiC+:\\
04573-1452   & 2250   & 5.0  & 1000  & 294  & \ 700  & 0.60  & 0.08 & SiC\\
05028+0106   & 2650   & 5.2  & 1000  & 332  & \ 700  & 0.10  & 0.12 & SiC+\\
05418-4628   & 2400   & 5.0  & 1000  & 500  & 1200   & 0.10  & 0.08 & SiC++:\\
05426+2040   & 2600   & 4.5  & 1000  & 554  & 1000   & 0.20  & 0.20 & SiC\\
05576+3940   & 2200   & 4.6  & 1000  & 403  & 1000   & 1.20  & 0.07 & Br2\\
06225+1445   & 2700   & 4.9  & 1000  & 400  & \ 850  & 0.03  & 0.12 & Br2\\
06331+3829   & 2550   & 3.7  & 1000  & 340  & \ 700  & 0.10  & 0.13 & SiC\\
06529+0626   & 2200   & 4.6  & 1000  & 282  & \ 700  & 1.00  & 0.07 & SiC \\
07045-0728   & 2400   & 6.5  & 1000  & 270  & \ 500  & 0.10  & 0.18 & SiC+\\
07057-1150   & 2650   & 3.0  & 1000  & 440  & \ 740  & 0.04  & 0.27 & SiC+:\\
07065-7256   & 2400   & 4.6  & 1000  & 282  & \ 700  & 2.20  & 0.07 & SiC\\
08538+2002   & 2400   & 5.0  & 1000  & 400  & 1200   & 0.20  & 0.04 & SiC++\\
09452+1330   & 2100   & 5.5  & 8000  & 110  & \ 500  & 10.0  & 0.01 & Red\\
10329-3918   & 2700   & 4.5  & 1000  & 390  & \ 850  & 0.04  & 0.11 & SiC+:\\
10350-1307   & 2700   & 4.5  & 1000  & 468  & \ 850  & 0.03  & 0.20 & SiC\\
10491-2059   & 2050   & 5.6  & 10000 & 270  & \ 950  & 0.70  & 0.02 & Red\\
12226+0102   & 2700   & 5.0  & 1000  & 487  & 1200   & 0.35  & 0.07 & Br1\\
12427+4542   & 2700   & 4.9  & 1000  & 317  & \ 700  & 0.05  & 0.10 & SiC+:\\
12447+0425   & 2200   & 4.3  & 1000  & 307  & \ 700  & 2.50  & 0.09 & SiC\\
12544+6615   & 2650   & 3.7  & 1000  & 155  & \ 400  & 0.04  & 0.06 & SiC+\\
15094-6953   & 2650   & 5.3  & 1000  & 416  & \ 700  & 0.03  & 0.26 & SiC+\\
18306+3657   & 2200   & 5.0  & 1000  & 317  & \ 700  & 0.03  & 0.10 & SiC++\\
19017-0545   & 2550   & 4.9  & 1000  & 315  & \ 800  & 0.10  & 0.07 & SiC+\\
19555+4407   & 2400   & 4.0  & 1000  & 761  & 1500   & 0.08  & 0.15 & SiC+:\\
21032-0024   & 2200   & 4.5  & 1000  & 294  & \ 700  & 2.50  & 0.08 & SiC \\
21399+3516   & 2800   & 4.0  & 10000 & 500  & 1000   & 0.04  & 0.14 & SiC+:\\
23587+6004   & 2500   & 3.0  & \ 800 & 554  & 1000   & 0.01  & 0.20 & Br1\\
       \hline
         \noalign{\smallskip} 
         \hline
      \end{tabular}
      \end{flushleft}
\end{table*}

\subsection{Results for SLMP stars}

{\bf {\it SiC} Class }

We analyzed 9 out of 40 SLMP objects in this class.  We found that  
effective temperatures vary between 2200 K and 2700 K.  Inner radii vary  
between 3.7R$_*$ and 5.0R$_*$, while most of the outer radii are 1000R$_*$. It 
must be kept in mind that the results are not very sensitive to this last 
parameter, as has been pointed out in previous works. The $\alpha$-SiC core 
(c$_{\rm SiC}$) grains vary between 280 to 550 \AA, and the amorphous carbon 
mantle (m$_{\rm A.C.}$) between 700 and 1050 \AA.  Optical depths for this 
class  vary between 0.10 and 2.5, with two exceptions: U Hya ($\tau$ = 0.03) 
and R For ($\tau$ = 7.0).  Finally SiC/A.C. ratios vary from 0.06 to 
0.20.  In Figures 1(a)-(d) we show out fits to RV Aqr and Y Tau. Solid 
lines represent the core/mantle grain model and dashed lines the two homogeneous 
grains model. Figures 1(b) and 1(d) show an enlarged view centered on the 
11.3$\mu$m feature. In all cases, best fits to this class were obtained using 
the two homogeneous grains model. 

\begin{figure}
\epsfxsize=\hsize
\epsfbox{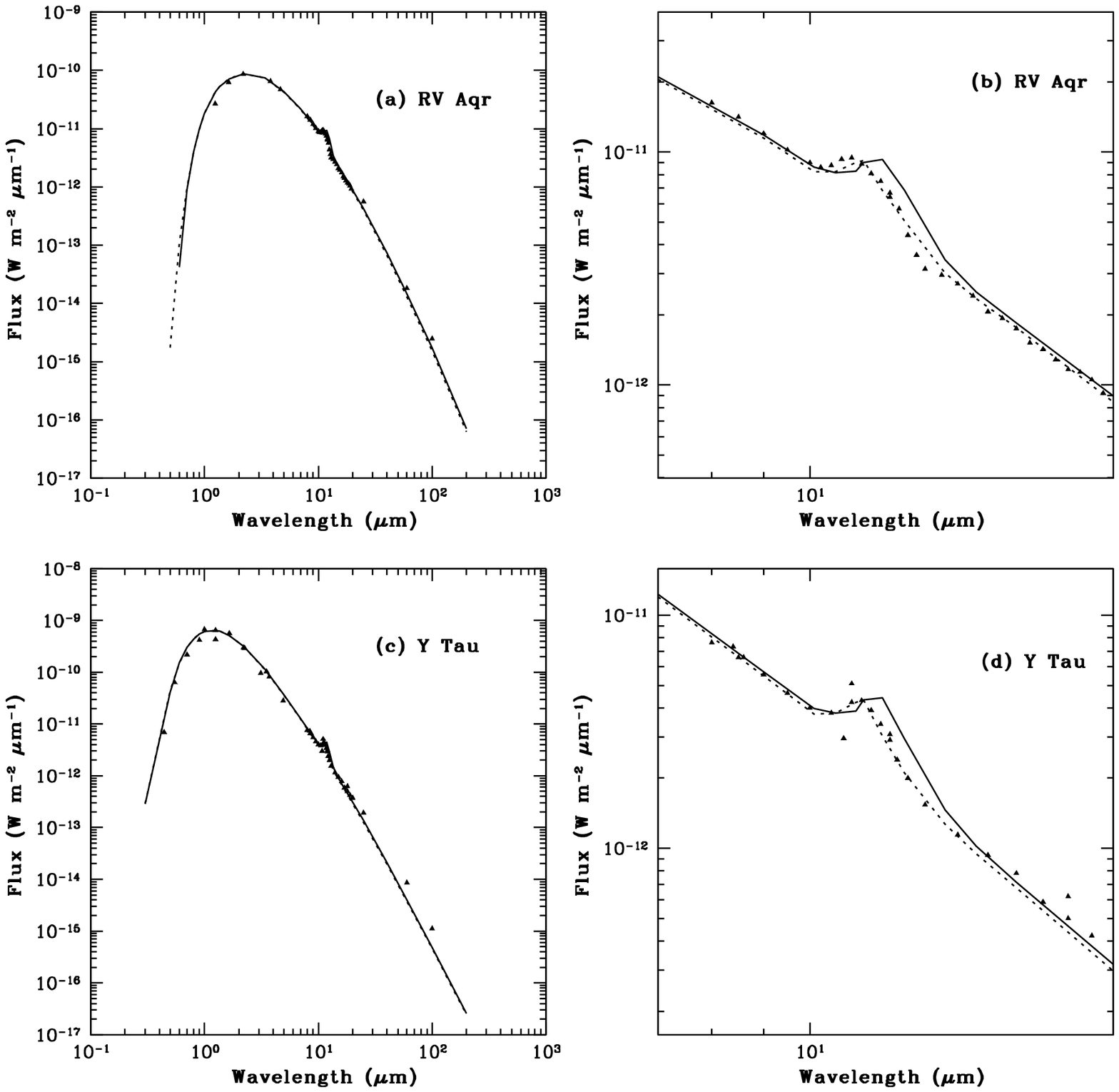}
\caption{Best models for {\it SiC} class.  In Figure 1(a)-(d) we plotted the 
best 
models using core/mantle  grains (solid line) and two homogeneous grain model  
(dashed line). Triangles represent the photometric data. In Figure 
1(b) and 1(d) we show an enlarged view of the 11.3 $\mu$m emission.}
\end{figure}

{\bf {\it SiC+} Class}

We have modeled 7 out 32 SLMP {\it SiC+} stars.  According to our results,  
the temperatures of central stars vary from 2400 K to 2650 K.  Outer dust 
envelope radii are the same for all stars (R$_{\rm 2}$ = 1000 R$_*$) and inner 
radii vary between 3.7 R$_*$ and 6.5 R$_*$. Sizes of mantle grains (m$_{\rm 
A.C.}$) show a great dispersion: 400 \AA\ to 1200 \AA.  The same occurs with 
the sizes of $\alpha$-SiC core (c$_{\rm SiC}$) which vary from 155 \AA\ to 500 
\AA. Optical depths have values between 0.03 and 0.50, and SiC/A.C. 
ratios between 0.06 and 0.26. Almost all stars were best described using the  
two homogeneous grains model; the unique exception is V Aql, which is better 
reproduced with a core/mantle grain code.  Figure 2 shows best fits to RY 
Mon and U Cam.

As a rule, we can say that the envelope of {\it SiC} and {\it SiC+} stars are 
nicely described by the existence of $\alpha$-SiC and A.C. homogeneous 
grains simultaneously.

\begin{figure}
\epsfxsize=\hsize
\epsfbox{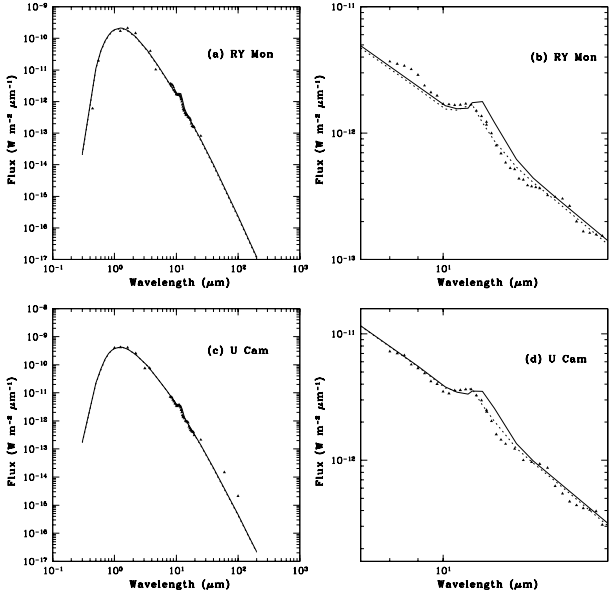}
\caption{This figure shows the best models for RY Mon and U Cam, which are in 
the {\it SiC+} class.  In Figure 2(a)-(d) we plotted models using core/mantle    
grain (solid line) and two homogeneous grain model (dashed line). 
Triangles represent the photometric data. In Figure 2(b) and 2(d) we show an 
enlarged view of the 11.3 $\mu$m emission.}
\end{figure}

{\bf {\it SiC++} Class}

We analyzed 5 out of 6 SLMP objects.  According to our models, effective 
temperatures are either 2200K or 2400K.  Almost all inner and outer radii have 
the same values: R$_1$ = 5R$_*$ and R$_2$ = 1000R$_*$. Mantle sizes 
(m$_{\rm A.C.}$) vary between 700 and 1200 \AA, and core sizes (c$_{\rm 
SiC}$) between 300 and 500 \AA. Optical depths vary between 0.02 and 0.20,  
which indicates very thin dust envelopes, and SiC/A.C. ratios between 0.04 and 
0.10.  Due to the absence of optical constants to describe the 8-9 $\mu$m and 
13 $\mu$m features, it is difficult to choose between the core/mantle grain or 
the two homogeneous grains model. We will discuss these results in the next 
section. Figure 3 shows two stars of this class.

\begin{figure}
\epsfxsize=\hsize
\epsfbox{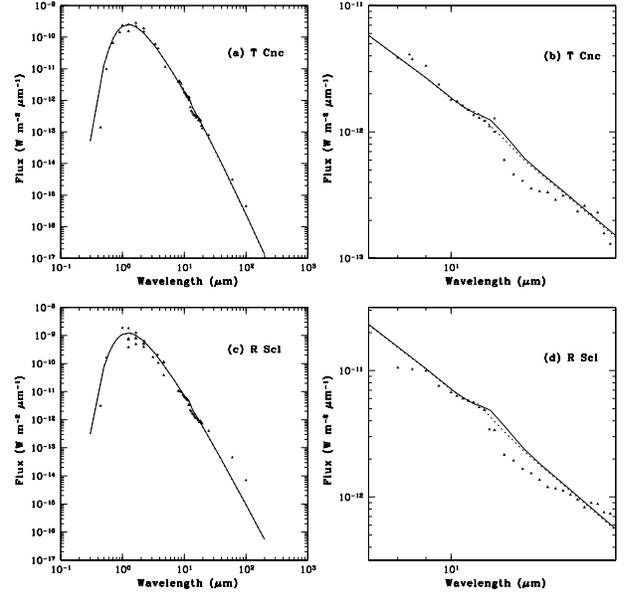}
\caption{This figure shows best models for TCnc and R Scl, which are in {\it 
SiC++} class.  In Figure 3(a)-(d) we plotted models using core/mantle  
grain (solid line) and two homogeneous grain model (dashed line). 
Triangles represent the photometric data. In Figure 3(b) and 3(d) we show an 
enlarged view of the 11.3 $\mu$m emission.}
\end{figure}

{\bf {\it Broad 1} and {\it Broad 2} Classes}

We have modeled 2 out 5 SLMP {\it Broad 1}, and 2 out 3 SLMP {\it Broad 2} 
stars. Our results indicate  very similar properties for both classes. The 
temperatures of the central stars are in the range of 2200K to 2700K.  Core and 
mantle sizes vary between 400-500 \AA\ and 850-1200 \AA, respectively.  They 
have thin envelopes with analogous dimensions. SiC/A.C. ratios varying between 
0.07 and 0.20 were found. We have obtained our best fits using a two homogeneous 
grains code for 3 stars. WZ Cas presents an absorption at about 14 $\mu$m, which 
is usually seen in J-type carbon stars, and it's difficult to distinguish 
between the models. SS Vir shows features at about 8 $\mu$m and 14 $\mu$m, too.   
Figure 4 presents best fit models to SS Vir ({\it Broad 1} class) and AZ Aur 
({\it Broad 2} class).  

\begin{figure}
\epsfxsize=\hsize
\epsfbox{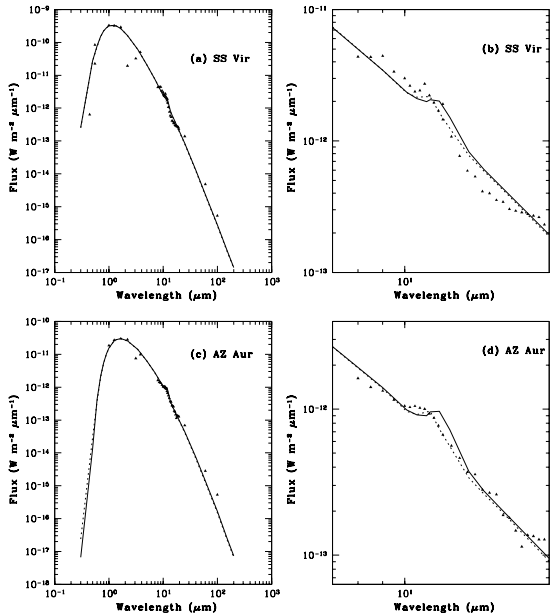}
\caption{This figure shows best models for SS Vir, which belongs to {\it Broad 
1} class and AZ Aur to {\it Broad 2} class.  In Figure 4(a)-(d) we plotted 
models using core/mantle  grain (solid line) and two homogeneous 
grain model (dashed line). Triangles represent the photometric data. In Figure 
4(b) and 4(d) we show an enlarged view of the 11.3 $\mu$m emission.}
\end{figure}

{\bf {\it Red} Class}

We have modeled 2 out of 3 SLMP stars: IRC+10216 and V Hya. Both objects 
are very well studied and supposed to have asymmetrical dust envelopes.  They 
are believed to be in the latest stages of stellar carbon evolution.  We will 
discuss them below.

{\bf (a) IRC+10216}

IRC+10216 has been extensively observed at optical, radio and infrared 
wavelengths. The central object is a long-period variable, with a period of 
640 days, and it is commonly considered to be a late-type carbon star.  Its 
envelope has been continuously surveyed and 380 molecular lines were detected, 
of which 317 have been identified (Cernicharo, Gu\'elin\& Kahane 2000). Deep B 
and V image-bands, reveal its extended circumstellar envelope in the dust 
scattered 
and show an episodic mass loss rate.  The circumstellar envelope is roughly 
spherically symmetrical but it is likely to be composed of a series of discrete, 
incomplete, concentric shells (Mauron \& Huggins 1999).

IRC+10216 was modeled by several authors.  Michell \& Robinson (1980) 
have used graphite grains and radiative transfer calculations were 
performed considering, as usual, a spherically symmetric envelope with a   
central star.  Rowan-Robinson \& Harris (1983),  Le Bertre (1987, 1988b), 
Martin \& Rogers (1987) and Griffin (1990) treated radiative transfer 
in the envelope of this star in a similar way.  Lorenz-Martins \& Lef\'evre 
(1993,1994) have modeled this star by considering a two homogeneous grain 
model consisting of $\alpha$-SiC and amorphous carbon grains 
simultaneously, as already cited. More recently, Groenewegen (1997) has computed 
a spherically symmetrical dust model and  suggested that IRC+10216 is in the 
latest phases of carbon stars evolution, like V Hya.

The effective temperature of IRC+10216 is 2100 K and our results lead to an 
extensive (R$_{\rm 1}$=5.6R$_*$ and R$_{\rm 2}$ = 8000R$_*$) but thick ($\tau$ = 
10) dust envelope.  The size of the amorphous carbon mantle grains (m$_{rm 
A.C.}$) was 500 \AA\ with an $\alpha$-SiC core of 110 \AA. Our best model was 
obtained considering core/mantle grains as can be seen in Figures 5(a) and 5(b).

{\bf (b) V Hya}

V Hya was classified as C6,5 by Yamashita (1972).  It is a variable 
 star with overlapping periods: a period of about 530 days with amplitude 
of about 1.5 mg and a longer period of 6500 days with amplitude of 3.5 mg.  This 
object is surrounded by an extended expanding molecular envelope, resulting from 
extensive mass-loss.  Mass-loss rate is in the range of 3.0 - 4.0 10 $^{-6}$ 
$M_{\odot}/yr$ (Knapp \& Morris 1985, Olofsson et al. 1990).  Polarimetric 
measures have been obtained by Johnson \& Jones (1991) and more recently by 
Trammell et al. (1994).  Johnson \& Jones (1991) have classified V Hya as a 
proto-planetary nebula and measured P(V) = 0.75 \% $\pm$ 0.02\% at 
$\theta$=21$^o$ $\pm$ 1$^o$.  Trammell et al. (1994) have observed this object 
in April 1992 and January 1993, and found that the polarization varied over this 
interval.  The envelope properties  found from molecular line observations 
by Knapp et al. (2000), like the fast molecular wind and the high mass loss 
rate, suggest that V Hya has entered its `superwind' phase. However, its 
spectral type, period, colors, and lack of ionizing radiation indicate that this  
star is still on the AGB. Then, V Hya is believed to be in the latest phases of 
mass loss on the AGB.  

Our results show that the temperature of the central star is 2050K. 
Contrary to what is expected for this evolved phase, we have found a thin 
($\tau$ = 0.7) and extensive dust envelope (R$_{\rm 1}$=5.5R$_*$ and R$_{\rm 2}$ 
= 10000R$_*$). The size of the amorphous carbon mantle (m$_{\rm A.C.}$) is 950 
\AA\ and the $\alpha$-SiC core is 270 \AA. SiC/A.C. ratio is very small, 
0.02. The best fit can be seen in Figures 5(c) and 5(d). We can say that a 
core/mantle  grain model fits this star well, but a two homogeneous grain 
model cannot be discarded.  

\begin{figure}
\epsfxsize=\hsize
\epsfbox{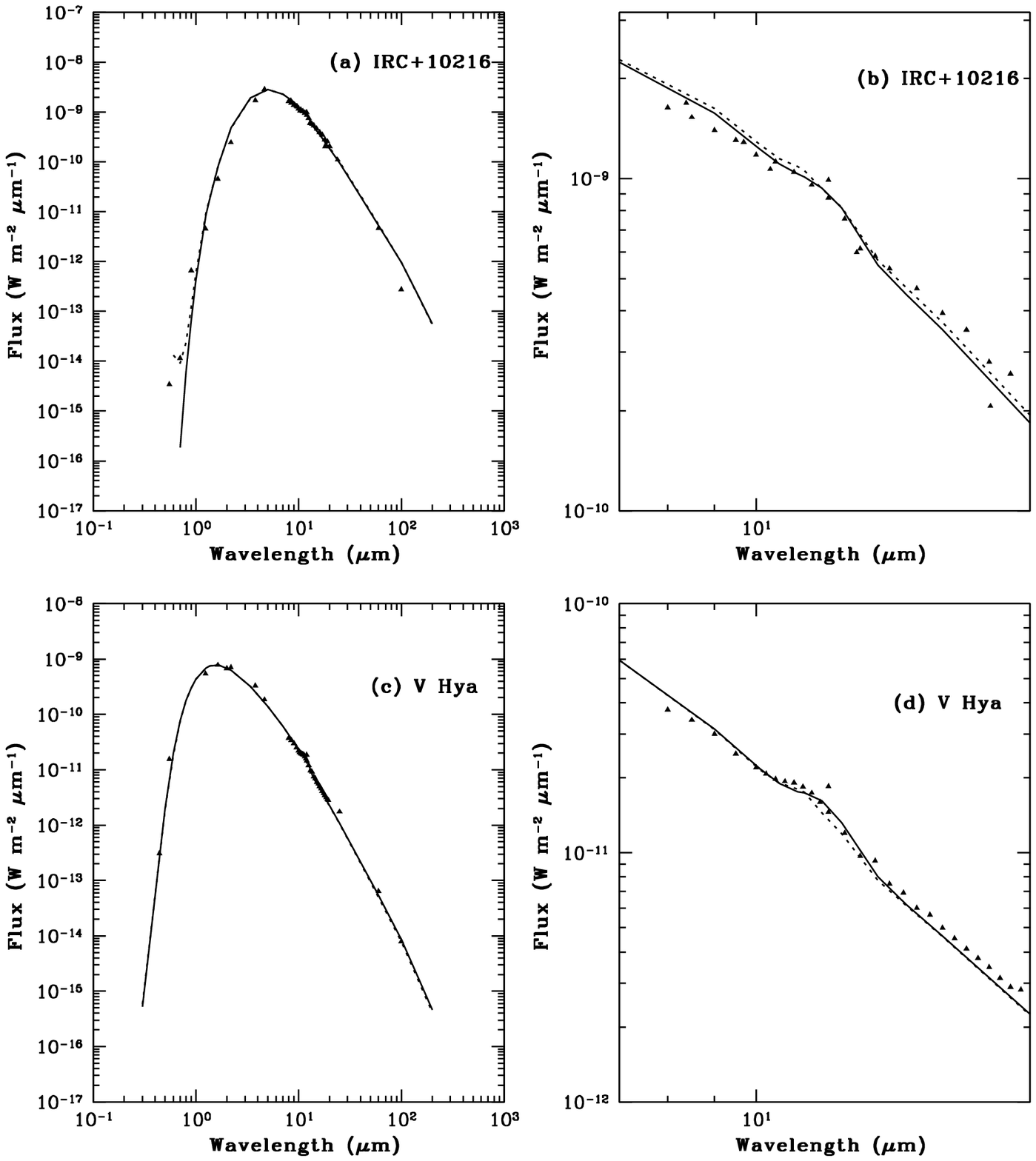}
\caption{This figure shows the best models for IRC+10216 and VHya which belong 
to {\it Red} class. In Figure 5(a)-(d) we plotted models using core/mantle  
grain (solid line) and two homogeneous grain model (dashed line). 
Triangles represent the photometric data.In Figure 5(b) and 5(d) we show an 
enlarged view of the 11.3 $\mu$m emission.}
\end{figure}

{\bf {\it SiC+:} Stars}

We analyzed 7 out 11 SLMP objects classified as {\it SiC+:}, where the 
colon means a more uncertain classification.  Almost all stars are well  
reproduced by a core/mantle grain model.  The only exception is AX Cyg, which 
seems to need single particles of $\alpha$-SiC, amorphous carbon and core/mantle 
grains simultaneously. The results obtained with the core/mantle grain code are 
similar to those found for {\it SiC+} class stars. The temperatures of the 
central stars vary from 2400K to 2800K. Inner radii vary between 3 to 4.9 
R$_{\star}$. Carbon mantle sizes (m$_{\rm A.C.}$) vary from 400\AA\ to 1500\AA\ 
with core (c$_{\rm SiC}$) values between 236\AA\ and 761\AA.  Optical depths are 
lower: 0.02 $\leq$ $\tau$ $\leq$ 0.08.On the other hand SiC/A.C. abundance 
ratios vary from 0.10 to 0.27. (For these stars we could say that the 
core/mantle grain models are more adequate, even with such small optical depths, 
since the SiC/A.C. ratios are higher.) Figure 6 shows an enlarged view of the 
two stars in this class, Y CVn and W CMa. We can see that the emission feature 
is shifted to longer wavelengths. This is a result of Mie's theory applied to 
core/mantle spheres. Suh (2000) has found a similar behavior in his models.

\begin{figure}
\epsfxsize=\hsize
\epsfbox{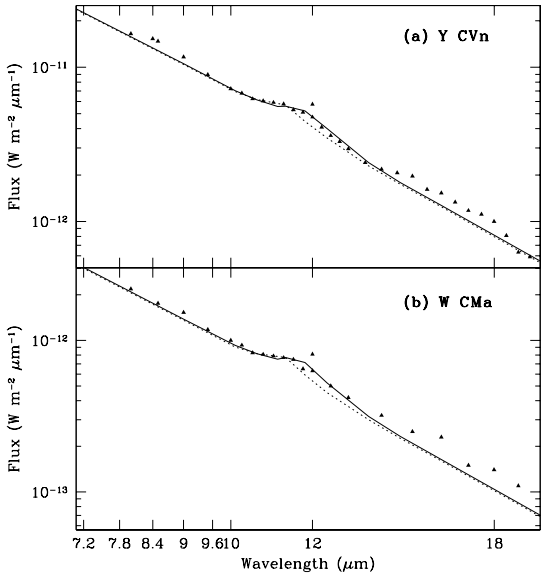}
\caption{This figure shows best models for Y Cvn and W CMa which belong to 
{\it SiC+:} class. In Figure 6(a-b) we plotted models using 
core/mantle  grain (solid line) and two homogeneous grain model (dashed line). 
Triangles represent the photometric data.}
\end{figure}

\subsection{Extreme Carbon stars}

  \begin{table*}
      \caption{results of the Sample with evolved stars}
      \begin{center}
      \begin{tabular}{llllllccr}
      \hline\noalign{\smallskip}
IRAS & $T_{\rm eff} (K) $  & R$_{\rm 1}$ (R$_{\star}$)  & R$_{\rm 1}$  
(R$_{\star}$) & $c_{SiC}$ (\AA) & $m_{A.C.}$ (\AA) & $\tau$ & SiC/AC &Env. 
Class\\ 
      \noalign{\smallskip}   
      \hline\noalign{\smallskip} 
05377+1346   & 2350  & 3.0  & 2000  & 282  & \ 700  & \ 3.0   & 0.07  & E1\\
05405+3240   & 2200  & 7.1  & 10000 & 215  & \ 700  & 10.6    & 0.03  & E2\\
06012+0726   & 2000  & 4.5  & 4000  & 189  & \ 900  & 12.5    & 0.01  & E2\\
06291+4319   & 2300  & 6.5  & 1000  & 226  & \ 700  & \ 3.5   & 0.04  & E1\\
06342+0328   & 2500  & 4.0  & 7000  & 256  & \ 700  & \ 7.0   & 0.05  & E2\\
07098-2112   & 2450  & 6.5  & 1000  & 236  & \ 700  & \ 4.0   & 0.04  & E1\\ 
08088-3243   & 2300  & 6.4  & 1000  & 215  & \ 700  & \ 4.0   & 0.03  & E1\\
15082-4808   & 2050  & 4.6  & 1000  & 273  & 1000   & \ 9.4   & 0.02  & E2\\ 
19594+4047   & 2000  & 5.0  & 7000  & 232  & \ 800  & 13.0    & 0.025 & E2\\ 
20570+2714   & 2200  & 8.0  & 1000  & 156  & \ 500  & \ 5.2   & 0.03  & E1\\ 
23257+1038   & 1900  & 7.0  & 1000  & 254  & \ 700  & 13.0    & 0.05  & E2\\  
\noalign{\smallskip} 
\hline
\end {tabular}
\end{center}
\end{table*}

Except for one case (R For), all {\it SiC} classes of the SLMP sample  contains 
carbon stars which have thin dust envelopes ($\tau$ $\leq$ 2.5). Our sample of 
extreme carbon stars contains objects that have optical depths varing between 
3.0 and 13. In order to make our analysis easier, we have decided to separate 
them according to this physical quantity: 3.0 $\leq$ $\tau$ $\leq$ 5.2 (E1 
group) and 7.0 $\leq$ $\tau$ $\leq$ 13.0 (E2 group).  

The stars belonging to our E1 group were better described by a two 
homogeneous grains model. Temperatures of the central stars are between 2200K 
and 2450K.  Inner radii are about 6.5 R$_*$ and outer radii 1000R$_*$ for all 
stars. Core (c$_{\rm SiC}$) and mantle (m$_{\rm A.C.}$) sizes are respectively 
about 200 \AA\ and 700 \AA\ in almost all cases.  The abundance ratios SiC/A.C. 
vary from 0.03 and 0.07. These results can indicate that such stars are related 
to the {\it SiC} SLMP class.  

On the contrary, stars belonging to our E2 group were better fitted with the 
core/mantle  grain code and are similar to the stars in the {\it Red} 
class.  They are cooler (1900K $\leq$ T$_{\rm eff}$ $\leq$ 2200K) and present a 
dust envelope more extensive than those of the E1 group.  Core and mantle sizes 
vary between 190 to 270 \AA\ and 700\AA\ to 1000\AA, respectively. The 
SiC/A.C. abundance ratios are also low, with values between 0.01 and 0.05.  In 
the E2 group, three stars were better represented by taking into account 
$\alpha$-SiC, amorphous carbon and core/mantle grains simultaneously. We can 
speculate that these three stars represent a transition phase between {\it SiC} 
and {\it Red} classes.  Figure 7 shows an enlarged view of AFGL 954 (E1 group) 
and AFGL 809 (E2 group).
 
\begin{figure}
\epsfxsize=\hsize
\epsfbox{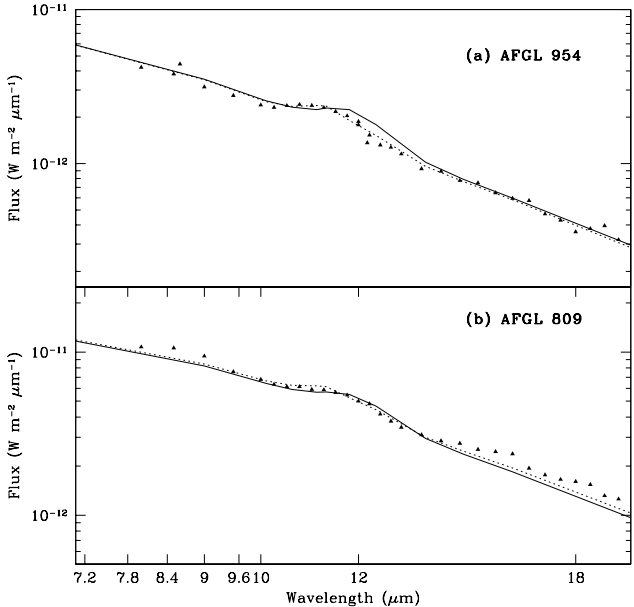}
\caption{This figure shows the best models for AFGL 954 and AFGL 809 which 
belong to E1 and E2 classes respectively. In Figure 7(a-b) we plotted models 
using core/mantle grain (solid line) and two homogeneous grain model (dashed 
line). Triangles represent the photometric data.}
\end{figure}
 
\section{Discussion and conclusions}

As commented in the previous section, the IR fluxes and the feature around
11.3 $\mu$m present in most sources that we have analyzed, are likely to be 
better reproduced using two homogeneous grains (A.C. and $\alpha$-SiC) 
simultaneously. This is true for {\it SiC}, {\it SiC+} and our {\it E1}
stars. On the other hand, a few objects (those in {\it Red} and in our {\it E2} 
class) are better described by a core/mantle  grain model. However, the   
{\it SiC ++} and {\it Broad 1} classes cannot be reproduced with the existing 
optical constants.
 
The {\it SiC++} class presents both the 8-9$\mu$m emission and also a very 
prominent 14$\mu$m absorption feature. The stars belonging to the {\it Broad 1} 
class also show a 14$\mu$m absorption. This absorption feature is weaker in 
Mira variables than in SR variables. This may be explained by a stronger dust 
emission in Mira variables which fills the molecular absorption (see Yamamura 
et al. 1998). Regarding the origin of the 8-9$\mu$m emission feature, Aoki, 
Tsuji \& Ohnaka (1999) have suggested that it may be a result of molecular 
absorption at 7.5$\mu$m and SiC emission at 11.3$\mu$m.  This absorption could 
be due to HCN and/or C$_2$H$_2$ photospheric absorption bands. On the other 
hand, the absorption feature at about 14$\mu$m was attributed by the same 
authors to HCN and C$_2$H$_2$ absorption in the photosphere or in the warm 
envelope close to the star.  This absorption feature would be formed in the 
inner envelope where the mid-infrared radiation originates. On the other hand, 
Yamamura et al. (1998) have studied the 14$\mu$m absorption in the ISO SWS 
spectra of 11 carbon stars with mass-loss ranging from 10$^{\rm -8}$ to 10$^{\rm 
-4}$ M$_{\odot}$/yr. According to these authors, all stars clearly show an 
absorption band at about 13.7$\mu$m due to C$_2$H$_2$ while the contribution 
from HCN molecules is small in this region.  

The stars of the {\it SiC++} class have thin envelopes that can favor 
the misidentification of spectral features like the 8-9$\mu$m one. On the other 
hand, this feature is also observed in some dust-enshrouded carbon stars, such  
as IRAS15194-5115, IRAS18239-0655 and IRAS18240+2326. Consequently, this feature 
could perhaps be produced by a solid component.  Indeed, Goebel, Cheeseman \& 
Gerbault (1995) have suggested $\alpha$:C-H, as described by Dischler, Bubenzer 
\& Koidl (1983).  Unfortunately, they have not published the set of optical 
constants to this component.  In order to try to reproduce the 8-9$\mu$m 
emission, we have modeled SS Vir, considering optical constants of the amorphous 
HAC as tabulated by Zubko (1996), but we have not been able to fit the emission. 
Bagnulo, Doyle \& Andretta (1998) have shown that this feature can be reproduced 
by a dust shell composed of a mixture of SiC grains and silicate grains. This 
interpretation, however, does not seem to be consistent with the theory of dust 
formation. Before discussing such results, let us comment now on some 
alternatives that might be investigated.

On the basis of a simple model of circumstellar envelopes, Kozasa et al. (1996) 
have proposed that the 11.3$\mu$m feature could be attributed to small spherical 
core/mantle  type grains (composed by a $\alpha$-SiC core and a carbon mantle) 
in most cases.  In their calculations they have used optical constants for 
SiC tabulated by Choyke \& Palick (1985), which peak at about 10.7$\mu$m. In 
fact, using this set of constants, the emission produced by a core/mantle  grain 
as proposed by them is about 11.3$\mu$m, and almost all sources of the SLMP 
sample could be fitted using a core/mantle  grain code. This is no longer true 
when we use the constants proposed by P\'egouri\'e (1988), which peak at about 
11.3$\mu$m.  In this case, the emission is shifted to longer wavelengths, too. 
We have computed our models considering the constants by P\'egouri\'e 
(1988) because the full treatment consisting of the Kramers-Kronig analysis has 
been taken into account.

Another possibility was raised by  Speck et al. (1999): they have fitted 
some carbon stars using $\beta$-SiC grains. Silicon carbide grains can be 
divided into two basic groups: $\alpha$-SiC if the structure is one of the many 
hexagonal or rhombohedral polytypes, and $\beta$-SiC if the structure is cubic. 
$\beta$-SiC feature occurs at about 0.4 $\mu$m shortwards of that of  
$\alpha$-SiC. Their results were obtained without the KBr correction and they 
determined that $\alpha$-SiC has an intense, broad band near 11.8$\mu$m and 
$\beta$-SiC peaks at 11.3 to 11.4$\mu$m.  Silicon carbide grains found in 
meteorites have isotopic compositions that imply that most of these grains were 
formed around carbon stars.  All studies to date of meteoritic SiC grains have 
found them to be of the $\beta$-type (Bernatowicz 1997). $\beta$-SiC will 
transform into $\alpha$-SiC above 2100$^{\rm o}$C but the reverse process is 
thermodynamically unlikely. The results obtained by Speck et al. (1999) show 
that there is an obvious predominance of the $\beta$-SiC phase and that there is 
now no evidence for the $\alpha$-SiC phase at all. Their sample contains {\it 
SiC}, {\it SiC+:}, {\it Red} and Extreme Carbon stars. However they do not solve 
the radiative transfer in these media. Moreover, we should expect some 
difference between ``early'' and ``late'' carbon stars with regard to dust 
grains. Unfortunately, the optical constants for $\beta$-SiC were calculated in 
a short range of wavelengths, about 7 to 12 $\mu$m. In this case, we need to 
adopt another set of optical constants at shorter wavelengths, where most of the 
stellar radiation is concentrated.  With this assumption we cannot prove that 
$\beta$-SiC grains are responsible for the 11.3 $\mu$m emission alone.

SLMP have suggested the following carbon-rich dust sequence: {\it SiC+} 
$\rightarrow$ {\it SiC} $\rightarrow$ {\it Red}.  The {\it Red} sources are 
significantly cooler on average than the {\it SiC+} sources.  Following our 
results, {\it SiC+} stars have thinner envelopes than {\it SiC} stars.  
Temperatures of the central stars are very similar but there is a tendency to 
cooler temperatures in this sequence.  Best models were obtained with 
two homogeneous grains for both {\it SiC+} and {\it SiC} class.  {\it Red} 
stars were best described with core/mantle  grains.  As mentioned before, 
based on our results we suggest that our sample of extreme carbon stars contains 
 {\it SiC} and {\it Red} stars.  In this sample, the thinner envelopes were best 
represented by two homogeneous grains models (our E1 group) while thicker ones 
by core/mantle  grain models (E2 group). The temperature of these stars are also 
cooler than in the {\it SiC+} class.  These results suggest that the sequence 
proposed by SLMP can be interpreted as an evolutionary scenario. Moreover, it 
seems reasonable to include our sample of extreme carbon stars in such a 
scenario. In the beginning of the sequence, grains are mainly composed of 
$\alpha$-SiC and amorphous carbon; with dust envelope evolution, carbon grains 
are coated in $\alpha$-SiC ones. (Hence the emission is shifted to longer 
wavelengths).  This phenomenon could perhaps explain the small quantities of SiC 
grains observed in the interstellar medium.

Concerning {\it SiC++} class stars, SLMP have proposed that they lie in a 
different evolutionary sequence, related to J-type carbon stars.  In their 
sample, which contains 96 objects, only nine are J-type  stars.  Two of them 
are classed in the {\it SiC++} class. One of us (Lorenz-Martins 1996) have 
proposed that J-type carbon stars have an alternative evolutionary scenario 
which differs from that proposed for ordinary carbon stars. In fact, according 
to the results of the present paper, {\it SiC++} stars have thicker (0.02 $\leq$ 
$\tau$ $\leq$ 0.20) envelopes than those of J-type stars (0.01 $\leq$ $\tau$ 
$\leq$ 0.05, see Lorenz-Martins, 1996).  It seems that some correlation between 
SiC/A.C. ratios of both groups of stars also exists. Such results reinforce the 
suggestion by SLMP linking {\it SiC++} and J-type carbon stars.  

\section*{Acknowledgments}
S.Lorenz Martins acknowledges FUJB (FUJB 8635-5) for financial support. We would 
like to thank Dr. R.Raba\c ca for his careful reading of the manuscript and also 
the referee, Dr.I. Little-Marenin for constructive comments and suggestions. 
This research was performed using the SIMBAD database at Strasbourg.

\end{document}